\documentclass[preprint,tightenlines,nofootinbib,floatfix]{revtex4}
\usepackage{graphicx,amsfonts,amssymb,amsmath}

\newcommand{\nc}{\newcommand}
\nc{\al}{\alpha}
\nc{\ga}{\gamma}
\nc{\de}{\delta}
\nc{\ep}{\epsilon}
\nc{\ze}{\zeta}
\nc{\et}{\eta}

\nc{\Th}{\Theta}
\nc{\ka}{\kappa}
\nc{\la}{\lambda}
\nc{\rh}{\rho}
\nc{\si}{\sigma}
\nc{\ta}{\tau}
\nc{\up}{\upsilon}
\nc{\ph}{\phi}
\nc{\ch}{\chi}
\nc{\ps}{\psi}
\nc{\om}{\omega}
\nc{\Ga}{\Gamma}
\nc{\De}{\Delta}
\nc{\La}{\Lambda}
\nc{\Si}{\Sigma}
\nc{\Up}{\Upsilon}
\nc{\Ph}{\Phi}
\nc{\Ps}{\Psi}
\nc{\Om}{\Omega}
\nc{\ptl}{\partial}
\nc{\del}{\nabla}
\nc{\be}{\begin{eqnarray}}
\nc{\ee}{\end{eqnarray}}
\nc{\ov}{\overline}
\nc{\gsl}{\not\!}

\newcommand{\bi}[1]{\bibitem{#1}}
\newcommand{\fr}[2]{\frac{#1}{#2}}

\newcommand{\GD}{\mbox{$\tilde{G}$}}
\newcommand{\gf}{\mbox{$\gamma_{5}$}}

\begin{document}
\preprint{TPI-MINN-99/34 $\;\;\;\;$ UMN-TH-1808-99}


\title{Theta Vacua, QCD Sum Rules, and the Neutron Electric Dipole Moment\\ $\;$\\}

\author{Maxim Pospelov\footnote{pospelov@mnhepw.hep.umn.edu}
        and Adam Ritz\footnote{aritz@mnhepw.hep.umn.edu}}

\affiliation{Theoretical Physics Institute, School of Physics and Astronomy \\
         University of Minnesota, 116 Church St., Minneapolis, MN 55455, USA\\$\;$\\ $\;$ \\ $\;$ \\}

\begin{abstract}
We present a detailed study of the electric dipole moment of the neutron
induced by a vacuum theta angle within the framework of QCD sum rules.
At next-to-next-to leading order in the operator product expansion, we find
the result $d_n(\theta) = 2.4\times 10^{-16}\bar \theta e\cdot cm$, 
to approximately
40\% precision. With the current experimental bound this translates
into a limit on the theta parameter of $|\bar \theta|<3\times 10^{-10}$.
We compare this result with the long-standing estimates obtained
within chiral perturbation theory, and observe a numerical similarity, but
also significant differences in the source of the dominant contribution.
\end{abstract}


\maketitle

\vfill\eject

\section{Introduction}

Electromagnetic observables which are odd under $T$ transformations
are an important source of information about CP properties 
of the physics at and above the electroweak scale, 
complementary to that coming from $K$ and $B$ meson physics. In 
particular, there are now impressive experimental limits on the 
electric dipole moments (EDMs) of 
neutrons, heavy atoms, and molecules \cite{nEDM,nEDM1,eEDM,mEDM,molEDM}. 
The Kobayashi-Maskawa model, so successful in explaining the observed
CP violation in K mesons, predicts EDMs to be several orders of 
magnitude smaller than the current experimental sensitivity. This
presents a unique opportunity for limiting extra sources of CP-violation, and
the constraints resulting from EDM data are generally 
very strong \cite{KL}. 

In principle, EDMs can be used to probe the physics at a
high energy scale by limiting the coefficients of operators
${\cal O}_i$ with dimension $k\geq 4$ in the effective low energy Lagrangian. 
The effective Lagrangian for these operators has the form,
\be
 {\cal L} & \sim & \sum_i c_i M^{4-k}{\cal O}_i^{(k)},
\ee
where $M$ is the mass scale at which these effective operators are induced
and $c_i$ their coefficients which, in general, 
have logarithmic scale dependence. 
These operators are odd under CP transformations and 
their coefficients $c_i$ are proportional to fundamental 
CP-violating phases of the underlying theory.

Of these contributions, the electron EDM operator is the only 
example which may be constrained while avoiding the uncertainties 
necessarily associated with strong interactions. 
For the EDM of the neutron and the $^{199}Hg$ atom, many more  
operators provide important contributions \cite{KKZ,KKY,W,KK,HP}. 
Particular operators of interest include schematically,
\be
 G\tilde{G},\;\;\frac{1}{M}\ov{q}F\si\ga_5 q, \;\;
 \frac{1}{M}\ov{q}G\si\ga_5 q, \;\;\frac{1}{M^2}GG\tilde{G},  
   \;\;\frac{1}{M^2}\ov{q}\Ga_1q\ov{q}\Ga_2 q,\;\;\mbox{ etc.},
     \label{ops}
\ee 
where $F$, $G$ and $q$ stand for electromagnetic, gluon and 
light quark fields, and 
$\Ga_{1,2}$ denotes various contributing matrix structures \cite{KKY}.  

In principle, the experiments \cite{nEDM,nEDM1,mEDM} 
impose strong constraints on the coefficients $c_i$. 
However, in practice, while
the operators can be perturbatively evolved down to a scale of order  
$1$ GeV, the ultimate connection between high energy parameters and 
low energy EDM observables necessarily involves non-perturbative physics.
It is this final link which we wish to consider in this paper.
The connection between different EDM observables and the coefficients $c_i$ 
are especially important in supersymmetric theories where the number 
of relevant operators is much smaller than in a generic scenario and 
$c_i$ can be explicitly calculated as a function of 
fundamental CP-violating SUSY phases
(For a recent discussion in the context of the MSSM, see e.g. \cite{FOPR}).

In the present paper,
we focus on the first of these contributions, $G\tilde{G}$. This
has a distinct status in that it has dimension=4 and thus receives 
contributions at tree-level from the fundamental QCD vacuum angle $\theta$,
the parameter labeling different super-selection sectors for the QCD vacuum.
Experimental tests of CP symmetry suggest
that $\theta$ is small and, among different CP-violating
observables, the EDM of the neutron ($d_n(\theta)$) is the most sensitive to its 
value \cite{nEDM,nEDM1}. However, the calculation of 
$d_n(\theta)$ is a long standing 
problem \cite{baluni,CDVW}. According to Ref.~\cite{CDVW}, 
an estimate can be obtained within chiral perturbation 
theory, relying on the numerical dominance of a one--loop diagram
proportional to $\ln{m_\pi}$ near the chiral limit. The result can be
conveniently expressed in the form
\be
d_n=e\theta\fr{m_u m_d}{f_\pi^2(m_u+m_d)}\left(\fr{0.9}{4\pi^2}
    \ln\frac{\Lambda}{m_\pi} + c\right)
\label{eq:log}
\ee
and is seemingly justified near the chiral limit where the 
logarithmic term may dominate over 
other possible contributions parametrized in this formula by the constant $c$.
However, these incalculable non-logarithmic contributions, can 
in principle be numerically more important than the logarithmic piece 
away from the chiral limit. In fact it is also worth noting
that in the limit $m_u,~m_d \rightarrow 0$, the logarithm is
still finite, and stabilized, for example, 
by the electromagnetic mass difference between the proton and neutron. 
Consequently, one is unable to estimate 
the uncertainty of the prediction \cite{CDVW}.
We note in passing that there is actually an additional $O(1)$ prefactor 
$(1-m_{\pi}^2/m_{\et}^2)$ associated with the vanishing of the $U(1)$
anomaly at large $N_c$, which will play a role subsequently. The explicit 
derivation of this factor will be given in this paper. 

If the logarithm is cut off at the neutron mass, $\Lambda \sim m_n$,  
and the non-logarithmic terms are 
ignored, one can derive a bound on the value of 
$\theta$ using the current experimental results on the EDM of 
the neutron \cite{nEDM,nEDM1}: $\bar\theta < O(10^{-10})$. 
Confronted with a naive expectation of $\theta\sim 1$, the 
experimental evidence for a small if not zero value for $\theta$ 
constitutes a serious fine tuning dilemma, usually referred to 
as the strong CP problem.  
As a consequence,
one is usually led to introduce some mechanism via which the primary
source of $\theta$, the fundamental vacuum angle, is removed. However,
even if this can be achieved, additional corrections are induced 
via the integration over heavy fields. Within this framework 
there are two main motivations
for refining the calculation of 
$d_n(\theta)$. 

The first refers to theories where the 
axion mechanism is absent and the $\theta$--parameter is zero at tree level 
as a result of exact P or CP symmetries \cite{P,CP}. 
At a certain mass scale these symmetries are 
spontaneously broken and a nonzero $\theta$ is induced through radiative 
corrections. At low energies, a radiatively induced theta term 
is the main source for the 
EDM of the neutron as other, higher dimensional, operators 
are negligibly small. As $\theta$ itself can be reliably calculated 
when the model is specified, the main uncertainty in predicting 
the EDM comes from the calculation of $d_n(\theta)$.

The second, and perhaps overriding, incentive to refine the calculation
of $d_n(\theta)$ is due to efforts to
limit CP-violating phases in supersymmetric theories in general, and
in the Minimal Supersymmetric Standard Model (MSSM) in particular. 
Substantial CP-violating SUSY phases contribute significantly to $\theta$ and 
therefore these models apparently require the existence of the
axion mechanism. However, as mentioned above, 
this does not mean that the $\theta$-parameter 
is identically zero. While removing $\theta\sim 1$, 
the axion vacuum will adjust itself to the minimum dictated by the 
presence of higher dimensional CP-violating operators which generate  
terms in the axionic potential linear in $\theta$. This induced
$\theta$--parameter is then given by:
\begin{eqnarray}
\theta_{induced}=-\fr{K_1}{|K|}, \;\;\mbox{where}\;\;  \label{eq:k1}
K_1=i\left\{\int dx e^{iqx}\langle 0|T(\fr{\al_s}{8\pi}G\GD(x),
{\cal O}_{CP}(0)|0 \rangle\right\}_{q=0}\nonumber,
\end{eqnarray} 
where ${\cal O}_{CP}(0)$ can be any CP-violating operator with dim$>$4 
composed from quark and gluon fields (as in (\ref{ops})), while 
\be
 K & = & i\left\{\int dx e^{iqx}\langle 0|T(\fr{\al_s}{8\pi}G\GD(x),
  \fr{\al_s}{8\pi}G\GD(0))|0 \rangle\right\}_{q=0} \label{K}
\ee
is the topological susceptibility correlator.
In the case of the MSSM, the most important 
operators of this kind are colour electric dipole moments of light quarks
$\bar{q}t^aG^a_{\mu\nu}\sigma_{\mu\nu}\gf q$, and three-gluon 
CP-violating operators. 
The topological susceptibility correlator $K$ was calculated in \cite{C,svz}
and the value of $\theta$ generated by color EDMs can be found in a 
similar way \cite{BUP}. Numerically, the contribution 
to the neutron EDM, arising from $\theta_{induced}$ is of the same order 
as direct contributions mediated by these 
operators and by the EDMs of quarks. Therefore, the complete calculation of 
$d_n$ as a function of the SUSY CP-violating phases must include  
a $d_n(\theta)$ contribution and a computation of this value, beyond the  
logarithmic estimate (\ref{eq:log}), is needed.  

In this paper, we present a detailed application of the 
QCD sum rule method \cite{sr} to obtain
an estimate for $d_n(\theta)$ beyond chiral perturbation theory.
Within currently available analytic techniques, QCD sum rules 
seems the most promising approach to this problem
as it has, in particular, been used 
successfully in the calculation of certain baryonic 
electromagnetic form factors \cite{is,BY}. 
Within the sum rule formalism, physical properties of the 
hadronic resonances are expressed via a combination of
perturbative and nonperturbative contributions, the latter parametrized in 
terms of vacuum quark and gluon condensates. We note that previously 
QCD sum rules were used to estimate the neutron EDM induced by a CP-odd color 
electric dipole moment of quarks \cite{KKY,kw}. Surprisingly, these results 
give $d_n$ $\sim 20$ times smaller than 
the estimates based on the chiral loop approach \cite{KK}. 
The calculation of $d_n(\theta)$ using QCD sum rules will certainly 
help to resolve this controversy. This question is of great numerical 
importance for the MSSM where color EDMs of quarks are large.

The approach we shall use follows recent work \cite{PR} on
the $\theta$-induced $\rho$--meson EDM in reducing the operator 
product expansion to a set of vacuum 
condensates taken in an electromagnetic and topologically nontrivial 
background. Expansion to first order in $\theta$ results in the appearance
of matrix elements which can be calculated via the use of current 
algebra \cite{C,svz}. In this approach the $\theta$--dependence 
naturally arises with the correct quark mass dependence, and the relation 
to the U(1) problem becomes explicit as $d_n(\theta)$ vanishes when the 
mass of the U(1) ``Goldstone boson'' is set equal to the mass of pion. 

The initial results from this study were presented in \cite{pr2}, and
in this paper we shall present the details of this analysis. We also
consider the relation of this result to the estimate (\ref{eq:log}) obtained
within chiral perturbation theory, and compare it with the outcome of 
an independent calculation of $d_n(\theta)$ reported recently in 
Ref.~\cite{CHM}. We begin in Section~II by studying
the phenomenological structure of the neutron correlator, and in
particular addressing the issue of how to ensure chiral invariance of the
result. In Section~III we perform a tree level OPE analysis of the
correlator to next-to-next-to leading order which corresponds to
sensitivity at the level of $O(1/q^2)$ terms. This requires an
investigation of mixing with an additional set of currents CP-conjugate
to the usual neutron interpolators. In Section~IV, we combine the
results of the previous two sections to construct a sum rule for
$d_n(\theta)$ which we analyze numerically and extract the estimate,
\be
d_n(\theta) & = & 2.4\pm 1.0\times 10^{-16}\bar \theta e\cdot cm.
\ee 
In section~V
we turn to chiral perturbation theory, and demonstrate 
explicitly the $m_\eta^2$--dependence of the EDM and the CP-odd
pion-nucleon coupling constant, indicating their explicit connection
to the $U(1)$ problem. We use this result to analyze and contrast
the large $N_c$ behavior of the chiral logarithm estimate and the QCD sum rule 
calculation of the EDM. We point out that, although unimportant
numerically, the two results differ in the large $N_c$ limit, with the 
chiral estimate suppressed by a relative factor of $O(1/N_c)$. We then 
conclude in Section~VI with some additional remarks.

\section{Phenomenological Parametrization and Chiral Invariance}

The starting point for the calculation is 
the correlator of currents $\et_n(x)$ with quantum
numbers of the neutron in a background with nonzero
$\theta$ and an electromagnetic field $F_{\mu\nu}$,
\be
 \Pi(Q^2) & = & i\int d^4x e^{iq\cdot x}
    \langle 0|T\{\et_n(x)\ov{\et}_n(0)\}|0\rangle_{\theta,F},
     \label{pi}
\ee
where $Q^2=-q^2$, with $q$ the current momentum.

Before turning to the OPE analysis of this correlator, it is convenient
to select an appropriate Lorentz structure to consider and, in the present
context, an important criterion will be invariance under chiral rotations.
It is crucial to consider this issue when CP-symmetry is 
broken by a generic quark-gluon CP-violating source -- 
the $\theta$--term in our case -- as 
the coupling between the physical state (neutron) described by a spinor $v$ 
and the current $\eta_n$ then acquires an additional phase factor
\be
\langle 0|\et_n|N\rangle=\lambda U_{\al}v, \;\;\;\;\;\;\;\;\;\;
     U_{\al}=e^{i\al\ga_5/2}. \label{coupling}
\ee
The existence of this unphysical phase $\al$ is already apparent when
one considers the sum rule for the neutron mass, which in the 
absence of CP-invariance can have an additional Dirac structure
proportional to $i\gamma_5$. When we turn to electromagnetic form factors, this
angle can mix electric ($d$) and magnetic ($\mu$) dipole moment structures 
and complicate the extraction of $d$ from the sum rule.

To see how this will work, we recall that when considering $\Pi$ in the
presence of some external field the phenomenological side of the sum rule 
may be parametrised by considering the form-factor Lagrangian which encodes 
the effective (in our case CP violating) vertices (see Fig.~1).
We recall that after expanding to leading order in 
the background field, $\Pi$ is effectively a three-point correlator, and
thus, although one can certainly write a two variable 
dispersion relation \cite{is}, it lacks the powerful 
positivity constraints which follow from the analytic structure of 
the two--point correlator. Therefore, its more appropriate to
instead explicitly parametrise $\Pi$ itself, rather than its
discontinuity.  
The corresponding form-factor Lagrangian
has the form ${\cal L}=\sum_n f_nS(q){\cal O}_nS(q)$, where
$f_n$ is the form factor, $S(q)$ is the on-shell propagator for the neutron
or one of its excited states, and ${\cal O}_n$ is the operator 
corresponding to the induced vertex.

Returning to the issue of chiral transformations, we can now
consider the effect of such a mapping on the the double pole contribution
on the phenomenological side of the sum rule. If we consider both
electric and magnetic dipole moments, then the double pole term
will have the form
\be
 \frac{P}{2(q^2-m_n^2)^2} & \equiv & \frac{1}{2(q^2-m_n^2)^2}
    (\gsl{q}+m_n)(\mu F\si-d\tilde F\si)(\gsl{q}+m_n).
\ee

\begin{figure}
 \includegraphics[width=5cm]{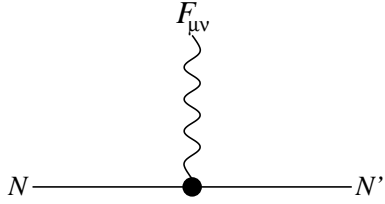}
 \caption{\footnotesize Hadronic contributions to the current correlator in an external
 electromagnetic field. Possible excited states with the neutron quantum
numbers are denoted generically by $N'$.}
\end{figure}

\noindent
in which we have introduced the dual field strength, 
$\tilde F_{\mu\nu}=\frac{1}{2}\ep_{\mu\nu\si\rh}F^{\si\rh}$.
Under a chiral rotation, we find that the numerator $P$ transforms as follows
\be
 U_{\al}PU_{\al} & = &
  m_n\{\gsl{q},\mu F\si-d\tilde F\si\}+
 m_n^2(\mu F\si-(d+\alpha \mu)(\tilde F\si)) \nonumber\\
 & & \;\;\;\;\;\;\;\;\;\;
 q^2(\mu F\si-(d-\alpha \mu)(\tilde F\si))+
 4q_\mu q_\nu\si_{\nu\lambda}
 (\mu F_{\mu\nu}-(d-\alpha\mu)\tilde F_{\mu\nu}),
\label{Lstr}
\ee
where we have retained only the zeroth 
and first order terms of the expansion in $\alpha$, and also neglected 
contributions proportional to $\alpha d$. 
We see that only Lorentz structures with an odd number of 
$\gamma$-matrices are independent of $\al$.  In calculating $d_n$, it is then
clear that we should study the operator $\{\tilde{F}\si,\gsl\!q\}$, 
as this is the unique choice with an unambiguous coefficient.

The phenomenological side of the sum rule will therefore be
parametrised in the form
\be
 \Pi^{(phen)} & = & \frac{1}{2}f(q^2)\{\tilde{F}\si,\gsl\!q\} +\cdots, 
          \label{phenfull}
\ee
where, since we work outside the dispersion relation we may add polynomials
in $q^2$ to ensure transversality in the chiral limit, 
and optimum behaviour for large $q^2$. We then find that the 
function $f(q^2)$ takes the usual form,
\be
 f(q^2) & = & \frac{\la^2d_nm_n}
     {(q^2-m_{n}^2)^2}+\sum_{i}\frac{f_i}{(q^2-m_{n}^2)
    (q^2-m_i^2)}+\sum_{i,j}\frac{f_{ij}}{(q^2-m_i^2)(q^2-m_j^2)},
     \label{f}
\ee
where $\la$ is the coupling of the current to the neutron state
(\ref{coupling}), $d_n$ is the neutron EDM, and $f_i$ and $f_{ij}$
correspond respectively to transitions between 
the neutron and excited states, and between
the excited states themselves. 

To suppress the contribution of excited states, we apply a Borel transform 
to $\Pi$, which we define, following \cite{svzrev,rry}, as
\be
 {\cal B}\Pi & \equiv & \mbox{lim}_{s,n\rightarrow\infty,s/n=M^2}
                    \frac{s^n}{(n-1)!}\left(-\frac{d}{ds}\right)^n\Pi(s),
\ee 
where $s=-q^2$. The continuum contributions in (\ref{f}) are then 
exponentially suppressed by a factor corresponding to the gap between
$m_n^2$ and the next excited state, usually taken around ($1.5$GeV)$^2$.
However, while this suppression is quite large, previous studies of 
CP-even sum rules have found that the continuum contribution is not
negligible (see e.g. \cite{rry}), and it is usually included 
for this reason. However, when
studying correlators in background fields, as mentioned above, one
is effectively dealing with a three-point function and one
consequently loses positivity constraints for the contributions. Thus
the couplings $f_{ij}$ for example are not sign--definite (as would be the
case for the two-point function). For this reason it seems inconsistent
to parametrise the continuum in the normal way, and we have no
alternative but to neglect it. In practice, we shall find 
that the sum rule we obtain is stable in any case. 

The coefficients of the single pole terms $f_i$ are also ambiguous in sign
for the same reason, but these contributions are not exponentially
suppressed by the Borel transform, and must be included for consistency.
We then find that the phenomenological expression takes the form
\be
 \Pi^{(phen)} &=& \frac{1}{2}\{\tilde{F}\si,\gsl q\}\left(
  \fr{\lambda^2d_nm_n}{(q^2-m_n^2)^2} +
   \fr{A}{q^2-m_n^2}+\cdots\right), 
          \label{phenfinal}
\ee
where the constant $A$ parametrizes all the single pole contributions and,
as we have explained, is not sign definite. It is this expression that
we shall contrast with the OPE calculation to be presented in the
next section.

\section{Calculation of the Wilson OPE Coefficients}

We now turn to a tree-level calculation of $\Pi$ (\ref{pi}) within the
framework of the operator product expansion. The neutron interpolating
current $\et_n$ is conveniently parametrised in the form,
\be
 \et_n & = & j_1 +\beta j_2,
\label{etan}
\ee
where the two contributions are given by
\be
 j_1 & = & 2\ep_{abc}(d_a^TC\ga_5u_b)d_c \\
 j_2 & = & 2\ep_{abc}(d_a^TCu_b)\ga_5d_c.
\ee

\begin{figure}
  \includegraphics[width=8.5cm]{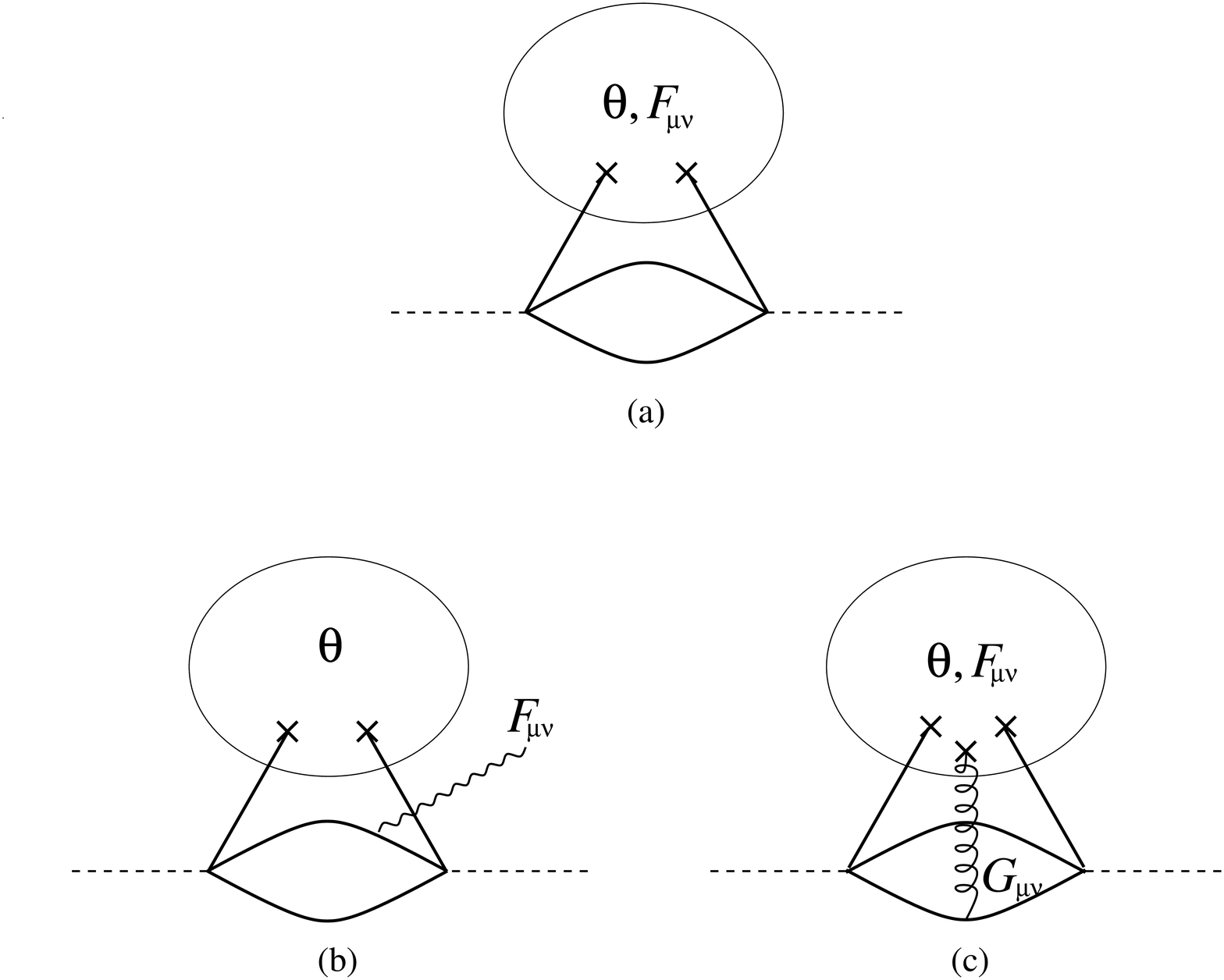}
 \caption{\footnotesize Various contributions to the CP-odd structure
  $\{\tilde{F}\si,\gsl{q}\}$. (a) is the leading order contribution
while (b) and (c) contribute at subleading order.}
\end{figure}

\noindent The current $j_2$ vanishes in the nonrelativistic limit, 
and lattice simulations have shown that $j_1$ indeed provides the 
dominant projection onto the neutron \cite{leinweber} 
(see also \cite{chung84}). 
From (\ref{coupling}), we may define 
the coupling to the neutron state in the form 
$\langle 0|\et_n|N\rangle=(\la_1+\beta\la_2)v$, where $\la_2\ll\la_1$.

Physical observables are independent of the parameter $\beta$, and 
in principle any combination of currents with neutron quantum numbers
is appropriate. However, within the sum rules formalism, one 
has the imperative of suppressing the contribution of excited states 
and higher dimensional operators in the OPE, and thus its convenient 
to choose $\beta$ to this end.  
Ioffe has argued \cite{ioffe81,bi83,ioffe83} 
that $\beta\sim -1$ is an apparently 
optimal choice from analysis of the the mass sum 
rule (see also \cite{leinweber}). An argument based on minimal sensitivity
\cite{chung82} leads instead to $\beta \sim -0.2$ using the same sum rule.
We shall return to this issue again later, 
but for the moment it will be convenient to keep $\beta$ arbitrary, and 
optimize once we have knowledge of the structure of the sum rule.
 
In the absence of CP violation, $j_1$ and $j_2$ form a basis for
projection onto the neutron state. However, the presence of a  
CP violating source means that, in principle, it's also necessary
to consider mixing with a CP-conjugate set of currents, which we
shall write as follows,
\be
  i_1 & = & 2\ep_{abc}(d_a^TCu_b)d_c \\
  i_2 & = & 2\ep_{abc}(d_a^T\ga_5 Cu_b)\ga_5d_c.
\ee
However, we shall show subsequently that it's actually possible to
choose a basis for $\theta$ in which these two sets of currents do not
mix, and the currents $(j_1,j_2)$ are a diagonalised combination for
projection onto the neutron state. Thus for the time being we shall
focus on $\et_n$ as the full current for calculation of $\Pi$.  

We now proceed to study the OPE associated with (\ref{pi}). 
The relevant diagrams we need to consider are shown in Fig.~2 
((a), (b) and (c)). 
In parametrizing $\theta$, we shall take a general initial condition
in which a chiral rotation has been used to
generate a $\ga_5$--mass, so that
\be
 {\cal L} & \sim & \cdots -\theta_q m_* \sum_f\ov{q}_fi\ga_5 q_f
       +\theta_G\frac{\al_s}{8\pi}G^a_{\mu\nu}\tilde{G}^a_{\mu\nu}+\cdots,
          \label{general}
\ee
in which we restrict to $q_f=u,d$, and so the reduced mass,
which plays an important role in CP-odd observables, has the form,
\be
 m_* & = & \frac{m_um_d}{m_u+m_d}.  \label{m*}
\ee
The physical $\theta$--parameter is of course $\overline{\theta}=\theta_q+\theta_G$, 
but we shall keep the general form (\ref{general}) and calculate the OPE 
as a function of {\em both} phases. The independence of the final answer
from $\theta_q-\theta_G$ will provide a nontrivial check on the consistency
of our approach. We shall find that this requires the consideration of 
mixing with the additional currents $(i_1,i_2)$, 
a point we shall come to shortly.

\subsection{Leading Order Contribution}

The leading order contribution is determined by the 1-loop diagram
in Fig.~2(a). We work as usual with a constant background electromagnetic
field, so that $A_{\mu}(x)=-\fr{1}{2}F_{\mu\nu}(0)x^{\nu}$, and
for later reference we also use a fixed point gauge
\cite{smilga} for the gluon potential,
$A_{\mu}^at^a(x)=-\fr{1}{2}G_{\mu\nu}^a(0)t^ax^{\nu}$. At leading order
it is not necessary to expand the quark propagator in the background
field, but it is necessary to consider the short distance
expansion of the quark wavefunction,
\be
 q(x) & = & q(0)+ x_{\al}D_{\al}q(0)+\cdots,
\ee
where $D_{\al}=\ptl_{\al}-ieA_{\al}$ is the covariant derivative in the
background field. When sandwiched between vacuum states, the second term
contributes at first order in the quark mass via use of the equation of
motion, $\gsl{\!D}q=-imq$. 
 
The vacuum structure is conveniently encoded in a generalized propagator
which incorporates the associated condensates. Projections onto
particular vacuum condensates are chosen in order to obtain the
Lorentz structure of interest (see e.g. \cite{PR} for more details
in the present context).
The electromagnetic field
dependence is determined in terms of the magnetic susceptibilities $\ch$,
$\ka$ and $\xi$, introduced in \cite{is}:
\be
 \langle 0| \ov{q}\si_{\mu\nu}q|0\rangle_F & = & \ch_q F_{\mu\nu}
           \langle 0| \ov{q}q|0\rangle \nonumber\\
  g\langle 0| \ov{q}(G_{\la\si}^at^a)q|0\rangle_F 
   & = & \ka_q F_{\mu\nu}\langle 0| \ov{q}q|0\rangle \label{susc}\\
 2g\langle 0| \ov{q}\ga_5(\tilde{G}_{\la\si}^at^a)q|0\rangle_F 
   & = & i\xi_q F_{\mu\nu}\langle 0| \ov{q}q|0\rangle \nonumber,
\ee
while the $\theta$--dependence is either explicit in the case of $\theta_q$,
or extracted via use of the anomalous
Ward identity (see e.g. \cite{svz}) in the case of $\theta_G$.
This use of the anomalous Ward identity was discussed for example
in \cite{PR}, and here we simply recall that the resulting
expression for a generic structure $m_q\langle\ov{q}\Ga q\rangle_{\theta_G}$
has the form,
\be
 m_q\langle0|\ov{q}\Ga q|0\rangle_{\theta_G} & = & im_*\theta_G
      \langle0|\ov{q}\Ga\ga_5 q|0\rangle+O(m_q^2),
\ee 
where the ability to neglect the $O(m_q^2)$ corrections 
follows since $m_{\et}\gg m_{\pi}$. The overall factor
has the form $(1-m_{\pi}^2/m_{\et}^2)$ \cite{svz} which vanishes 
when $U(1)$-symmetry is restored ($m_\eta\rightarrow m_\pi$) \cite{PR}.
Making use of the anomaly, we see that the result 
then has formally the same form as arises
from the $\theta_q$ $\ga_5$--mass contribution. This is once again
a consequence of the anomaly, but an important point is that the
sign of these contributions may differ, and thus we may obtain
contributions having the unphysical form $\theta_G-\theta_q$. The
resolution of this puzzle will be described shortly. 

Defining $iS(q)\equiv \langle 0|q_a(x) \ov{q}_b(0)|0\rangle_{F,\theta}$,
and ignoring a trivial $\de$--function over colour indices, the 
leading order propagator adapted to the CP-odd sector
and appropriate for Fig.~2(a), then takes the form,
\be
 S_{LO}(x) & = & \frac{\gsl x_{ab}}{2\pi^2x^4}
       +\frac{im_*}{4\pi^2x^2}(1-i\theta_q\ga_5)_{ab} \nonumber \\
    && \!\!\!\!\!\!\!\!\!\!\!\!\!\!\!\!\!\!\!\! 
   -\frac{\ch_qm_*\ov{\theta}}{24}\langle \ov{q}q\rangle F_{\al\beta}x_{\al}
   (\ga_{\beta}\ga_5)_{ba}
       +\frac{i\ch_q}{24}\langle \ov{q}q\rangle(F\si(1+i\theta_G\ga_5))_{ba}, 
      \label{prop}
\ee
We shall henceforth follow \cite{is} and assume that $\ch_q=\ch e_q$ etc.,
with flavour independent parameters $\ch,\ka,\xi$. 

Substituting the generalized propagator into (\ref{pi}) according to
the allowed contractions, performing the rather lengthy but
straightforward algebraic manipulations, and Fourier transforming
to momentum space, we find the result,
\be
 \Pi_{LO}(q^2) & = & -\frac{\ch m_*}{64\pi^2}\langle \ov{q}q\rangle
      \{\tilde{F}\si,\gsl q\}
   \left[\ov{\theta}(\beta+1)^2(4e_d-e_u)
   +\tilde{\theta}(1-\beta^2)(2e_d-e_u)\right]\ln \frac{\La^2}{-q^2},\;\;\;\;
\label{opejj}
\ee
where $\ov{\theta}=\theta_q+\theta_G$, and $\tilde{\theta}=\theta_G-\theta_q$ is an
unphysical combination. The appearance of this unphysical
combination is somewhat surprising and, as one might anticipate, is
due to an additional source of mixing. The additional currents
one needs to consider are precisely the CP conjugate set $(i_1,i_2)$
introduced earlier on. 

In order to illustrate the effect of this mixing we shall consider
first the OPE for the two-point functions 
\be
\langle 0| T\{j_{1}, \bar j_{1}\}| 0 \rangle , ~ 
\langle 0| T\{i_{1}, \bar i_{1}\}| 0 \rangle , ~ 
\langle 0| T\{j_{1}, \bar i_{1} \}| 0 \rangle , ~
\langle 0| T\{i_{1}, \bar j_{1}\}| 0 \rangle.
\label{ij}
\ee
and the Lorentz structure proportional to $\gsl q$. When both phases are set 
to zero, $\theta_G=\theta_q=0$, only the diagonal, $ii$ and $jj$, 
two-point functions survive and the cross--terms in (\ref{ij}) are 
identically zero. When the $\theta$-phases are not zero, the 
$ij$ correlators no longer 
vanish and an explicit calculation gives,
\be
 i\int d^4x e^{iq\cdot x}\langle 0| T\{j_{1}, \bar i_{1} \}| 0 \rangle =
i(\theta_G-\theta_q)\fr{3}{4\pi^2}m_*\langle \bar qq\rangle \gsl q
\ln \frac{\La^2}{-q^2}\\
 i\int d^4x e^{iq\cdot x}\left(\langle 0| T\{j_{1}, \bar j_{1} \}| 0 \rangle-
\langle 0| T\{i_{1}, \bar i_{1} \}| 0 \rangle\right )= 
\fr{3}{2\pi^2}m_*\langle \bar qq\rangle \gsl q\ln \frac{\La^2}{-q^2}. 
\nonumber
\end{eqnarray}
A straightforward re-diagonalization produces two ``eigencurrents'', $j_1 + 
\fr{i\tilde \theta}{2}i_1$ and  $i_1 - \fr{i\tilde \theta}{2}j_1$,
not mixed in the presence of $\theta_G$ and $\theta_q$. 
A similar procedure can be performed for the currents $j_2$ and $i_2$.
The mixing between  $j_1$ and  $i_2$,  and $j_2$ and $i_1$, is absent even
for nonzero $\theta_G,~\theta_q$ because these currents  differ by 
an overall $\gamma_5$ matrix which effectively gives an anticommutator 
with $\gsl{\!q}$.
Therefore, the generalization of the neutron interpolating current 
(\ref{etan})  can be written in the following form,
\be
\eta_n = j_1 + \beta j_2 + \fr{i\tilde \theta}{2}\left( i_1 +\beta i_2 \right)
\label{etan1}
\ee
Since the mixing between these two sets is explicitly proportional to 
$\theta_G-\theta_q$, it is convenient to take $\theta_q=\theta_G$ as a useful choice 
of basis when working with $j_1$ and $j_2$, where the mixing between 
$j_{1,2}$ and $i_{1,2}$, is simply absent. This situation resembles 
the problem in obtaining the ``correct'' quark mass behavior for 
the EDM of $\rho$, addressed in \cite{PR}, in which one can simplify
calculations by choosing $m_u=m_d$. Alternatively, one can 
use the generalized form for the current and observe 
that the presence of extra terms in 
(\ref{etan1}) gives a contribution to the OPE cancelling identically 
the $\tilde \theta$--dependence of eq. (\ref{opejj}).

The outcome, either by including the mixing terms with $(i_1,i_2)$, or
by choosing the basis $\theta_q=\theta_G$, is the same, and we find
\be
 \Pi_{LO}(q^2) & = & -\frac{\ch m_*}{64\pi^2}\langle \ov{q}q\rangle
      \{\tilde{F}\si,\gsl q\}
   \left[\ov{\theta}(\beta+1)^2(4e_d-e_u)\right]\ln \frac{\La^2}{-q^2},
\ee
explicitly proportional to the physical combination $\ov{\theta}$.

\subsection{Subleading Contributions}

Diagrams (b) and (c) in Fig.~2
require, in addition, the leading order expansion in the background gluon and
electromagnetic fields, and $S_{NLO}$ is consequently more involved.
We concentrate first on Fig.~2(b), which actually does not provide many
non-zero contributions. In addition to the leading order
propagator presented in (\ref{prop}), the expansion in the
background field leads to,
\be
 S_{NLO}^F & = & \frac{e_q}{8\pi^2}\frac{x_{\al}}{x^2}\tilde{F}_{\al\beta}
      (\ga_{\beta}\ga_5)_{ab} 
      - \frac{im_qe_q}{32\pi^2}\ln(-x^2)(F\si(1-i\theta_q\ga_5))_{ab} \nonumber\\
   & & \;\;\;\;\;\; +\frac{i}{12}\langle \ov{q}q\rangle(1+i\theta_G\ga_5)_{ab}.
\ee
With the field strength appearing explicitly, it is not necessary 
to expand the quark wavefunction, thus simplifying the calculation.
The resulting expression for $\Pi$ actually receives no contributions
from the first term of $S_{NLO}$ as it leads to a different Lorentz 
structure. However, we see that since the
propagator is logarithmic in the second term, 
the corresponding result behaves like $\ln(-x^2)/x^4$ which
gives an infrared divergence in the Fourier transform to momentum 
space. These divergences were also observed in \cite{is}, and we 
cut it off at a scale $\mu_{IR}$. Strictly, the full OPE should be
independent of such a cutoff, this dependence in the coefficient
being cancelled by a similar dependence in the condensates. We shall ignore
this subtlety here as these logarithmic terms will not enter our
final sum rule. With this caveat, the resulting
contribution to the $\{\tilde{F}\si,\gsl{q}\}$ structure in 
$\Pi$ has the form,
\be
 \Pi(Q^2)_{NLO}^{(b)} & = &  \frac{\ov{\theta}m_*}{16\pi^2}\langle 
    \ov{q}q\rangle\{\tilde{F}\si,\gsl q\}
   (\beta-1)^2e_d\left(\ln\frac{Q^2}{\mu_{IR}^2}-1\right)\frac{1}{Q^2},
\ee
where we have checked that on calculating the mixing with $(i_1,i_2)$
the unphysical dependence on $\theta_G-\theta_q$ drops out and we are
left with the expression presented here.

Diagram (c) in Fig.~2 requires considerably more work, as there are
several classes of contributions arising from it. Firstly, we need
to expand the quark propagator in the background gluon field, which
gives rise to terms analogous to those for $S_{NLO}^F$ given above, since
we make use of a fixed point gauge. Secondly, it is also possible to
project such terms onto the vacuum in order to extract the leading
dependence on $F_{\mu\nu}$ using (\ref{susc}). This requires
a first order expansion of the quark wavefunction\footnote{\small Note also
that a second order expansion of the wavefunction, making use of the 
leading order propagator, would also provide contributions at
subleading order, but these are combinatorially highly suppressed \cite{is}
relative to the terms we are considering, and so will be ignored.}.
The resulting propagator, ignoring contributions which lead to
other Lorentz structures, may be written in the form
\be
 S_{NLO}^G & = & S_{(1)}^G+S_{(2)}^G,
\ee
where $S_{(1)}^G$ is the expansion in the background gluon field,
\be
 S_{(1)}^G & = & \frac{g}{8\pi^2}\frac{x_{\al}}{x^2}\tilde{G}_{\al\beta}
      (\ga_{\beta}\ga_5)_{ab}
      -\frac{im_qg}{32\pi^2}\ln(-x^2)(G\si(1-i\theta_q\ga_5))_{ab},
\ee
while $S_{(2)}^G$ conveniently encodes the dependence on the condensates
in the form,
\be
 G_{\al\beta}S_{(2)}^G & = & 
      \frac{i}{32}m_*\ov{\theta}\xi_q\langle \ov{q}q\rangle
       x_{\rh}F_{\al\beta}(\ga_{\rh})_{ab}
    -\frac{i}{24}\langle \ov{q}q\rangle(i\xi_qF_{\al\beta}\ga_5+2\ka_q
        \tilde{F}_{\al\beta})(1+i\theta_G\ga_5)_{ab}.
\ee
In calculating the corresponding contributions to $\Pi$, it is understood
that one picks out the appropriate cross-terms in the correlator, in order
to extract the leading order dependence on $\theta$, $m_*$ and $F_{\mu\nu}$.
Infrared divergent logarithms also arise in this case from the
final term in the propagator, while the other contributions actually
contribute at next-to-next-to leading order, which in this case is
$O(1/Q^2)$. After a lengthy calculation, we obtain the following
contributions to $\Pi$,
\be
 \Pi(Q^2)_{NLO}^{(c)} & = & \frac{\ov{\theta}m_*}{64\pi^2}\langle \ov{q}q\rangle
      \{\tilde{F}\si,\gsl q\}
   \left[(\beta-1)^2e_d(2\ka+\xi)\left(\ln\frac{Q^2}{\mu_{IR}^2}
       -1\right)\frac{1}{Q^2}\right. \nonumber \\
   &&\!\!\!\!\!\!\!\!\!\!\!\!\!\!\!\!\!\!\!\!\!\!\!\!\!\left.
   \frac{\xi}{2}\left((4\beta^2-4\beta+2)e_d+(3\beta^2+2\beta+1)e_u
     \right)\frac{1}{Q^2}\cdots\right],
\ee
where we have checked that the unphysical $\theta_G-\theta_q$ logarithmic
terms are cancelled by mixing with $(i_1,i_2)$, while we have
simply used the ``gauge'' $\theta_G=\theta_q$ to evaluate the $O(1/Q^2)$ terms
as these will turn out in fact to be numerically insignificant. 

Putting all the pieces together, we
present the final result for the OPE structure arising from
diagrams (a), (b), and (c) in Fig.~2,
\be
 \Pi(Q^2) & = & -\frac{\ov{\theta}m_*}{64\pi^2}\langle \ov{q}q\rangle
      \{\tilde{F}\si,\gsl q\}
   \left[\ch(\beta+1)^2(4e_d-e_u)\ln \frac{\La^2}{Q^2}\right. \nonumber \\
   &&  
    -4(\beta-1)^2e_d\left(1+\frac{1}{4}(2\ka+\xi)\right)
     \left(\ln\frac{Q^2}{\mu_{IR}^2}
       -1\right)\frac{1}{Q^2} \nonumber \\
   && \left.
  - \frac{\xi}{2}\left((4\beta^2-4\beta+2)e_d+(3\beta^2+2\beta+1)e_u
     \right)\frac{1}{Q^2}\cdots\right].
\label{ope}
\ee
This is our final result for the OPE, and we shall
analyze the resulting sum rule obtained by equating this result with the
phenomenological parametrization discussed in Section 2, in the next
section.

However, before turning to this analysis, we shall first make some
comments regarding additional contributions that we have ignored.
Some additional classes of diagrams are shown in Fig.~3.

\begin{figure}
 \includegraphics[width=8.5cm]{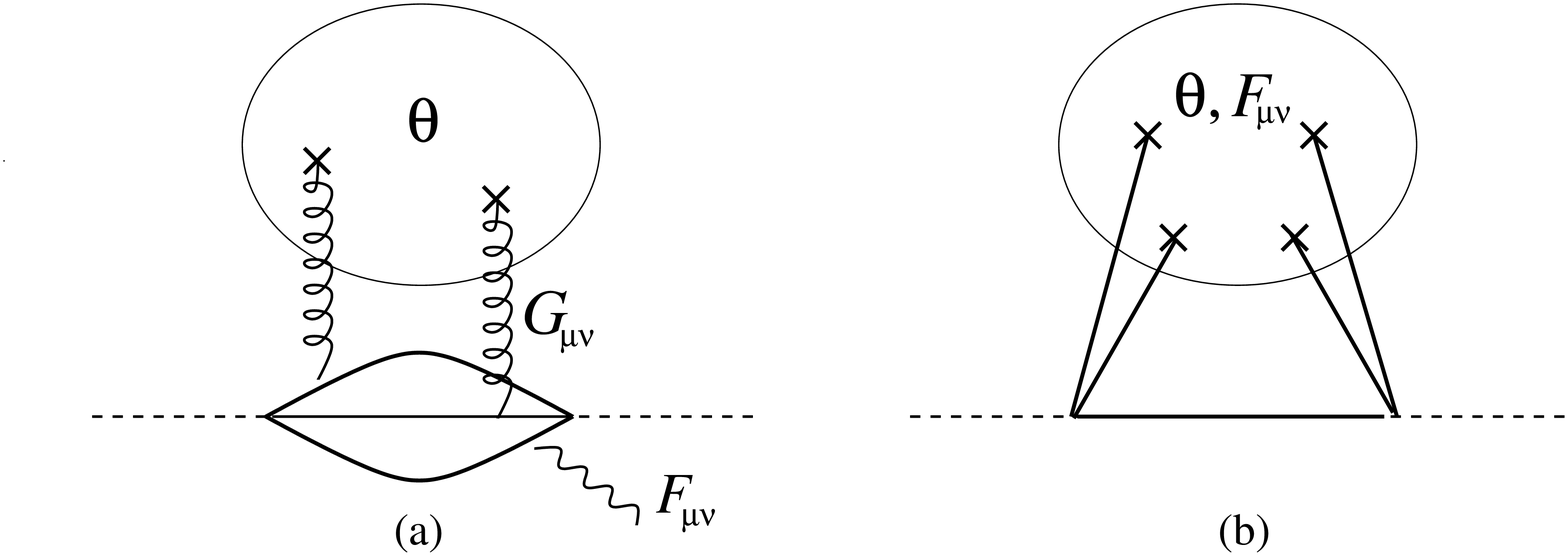}
 \caption{\footnotesize Additional subleading contributions to the OPE.}
\end{figure}

\noindent Naively, the diagram in Fig.~3(a) is loop suppressed relative
to the contributions we have considered. This loop suppression is, however,
fictitious as this diagram is proportional to the vacuum correlator 
$\alpha_s(4\pi)^{-1}\int d^4 x \langle 0|(G\tilde G),(G\tilde G)|0 \rangle$
which is the same as 
$\int d^4 x \langle 0|(G\tilde G), m_*\ov{q}q|0\rangle$.
Nonetheless, we find in practice that all such contributions vanish.

Diagrams of the form shown in Fig.~3(b) are more problematic
because they involve correlators of the form 
\be
\Pi_{3(b)}\sim  
\int d^4x e^{iqx}
    \langle 0|T\{(\bar qq)^2(x),m_*\sum_f\ov{q}_fi\ga_5 q_f
 \}|0\rangle. \!\!
\ee
which are not calculable within our chiral approach.
However, one suspects that these contributions of
$O(\langle \ov{q}q\rangle^2)$, although suffering
no loop-factor suppression, are small due in part
to the small numerical size of $\langle \ov{q}q\rangle$,
but also due to combinatorial factors. 
We estimate these contributions via saturation with the physical 
$\eta$ meson. The result indeed turns out to be parametrically 
smaller than any term listed in Eq.~(\ref{ope}). 

We shall conclude this section with some brief remarks on an
independent calculation of $d_n(\theta)$ using QCD sum rules,
recently reported in \cite{CHM}. The approach taken in this work
is somewhat complementary to ours, as the authors use different
Lorentz structures in (\ref{Lstr}). These structures are not chirally 
invariant and have an admixture of the magnetic moment $\alpha\mu$. 
Consequently, this requires simultaneous treatment of various two- 
and three-point correlation functions. The authors of \cite{CHM}
introduce a nice chirally covariant notation to assist in keeping
track of these terms. In their approach, one must then try to isolate the
contributions to $d_n$ within sum rules that depend also on the neutron
mass, $\mu$, and the phase $\al$. The subtlety here is that one must
carefully keep track of all terms of $O(m_q)$, despite the fact that 
$\mu$ for example is determined to leading order at $O(m_q^0)$.
As a result, the separation of $d\sim m\theta\langle \bar qq\rangle$ 
arises as a delicate cancelation between a combination of different terms, 
each of order $O(m_q^0)$. The authors of \cite{CHM} simplify matters
somewhat by setting all quark masses equal $m_u=m_d$, and therefore
don't observe the appearance of the reduced mass $m_*$ (\ref{m*}).

At first sight there is  
a significant problem with the use of Lorentz structures other than 
$\{\tilde{F}\si,\gsl{\!q}\}$, which is the appearance at leading order of 
certain incalculable condensates of the form
\be 
    \langle 0|T\{\bar qi\gf q(x),m_*\sum_f\ov{q}_fi\ga_5 q_f
 \}|0\rangle\;\;\;\;{\rm and}\;\;\;\;
    \langle 0|T\{\bar q\si_{\mu\nu}q(x),m_*\sum_f\ov{q}_fi\ga_5 q_f
 \}|0\rangle. \!\!
\label{noncl}
\ee
In our approach these terms arise manifestly
at $O(m_q^2)$, while for other channels these terms can arise at $O(m_q)$
and apparently need to be dealt with. The first condensate 
in (\ref{noncl}) can be connected to an $O(m_*)$ correction to the correlator 
$\int d^4x\langle 0|T\{\bar qi\gf q(x), (G\tilde G)(0)\}|0\rangle$,
for which no reliable means of extracting its value is known (for a 
recent discussion of the subleading mass dependence of this and related 
correlators see, e.g. \cite{Smilga}).
These terms, as well as $\xi$ and $\kappa$--proportional contributions, 
are ignored in \cite{CHM}. However, this is consistent as 
one may show that, remarkably enough, 
in the final result for the ratio $d_n/\mu_n$ these corrections
only appear at $O(m_q \langle \ov{q}q\rangle^2)$; a subleading effect.
This wasn't explicitly pointed out in \cite{CHM}, but we believe it is an 
important (and indeed interesting) point, 
necessary for the consistency of their approach.
Numerically, the results of \cite{CHM} are very close to the
results we obtained in \cite{pr2}, and shall discuss in the next section.
Thus, the two approaches indeed appear quite complementary.

As a final related comment, we note that
the use of chirally non-invariant channels may be the origin of 
numerical problems in the QCD sum rule calculations of $d_n$ 
induced by color EDMs \cite{KKY,kw}. This, of course, is yet to be checked but
our preliminary estimates show that the use of the chirally invariant 
channel yields the neutron EDM at a level comparable with phenomenological 
estimates \cite{KK}.

\section{Numerical Analysis of the Sum Rule}

Turning now to the analysis of the sum rule (\ref{ope}),
an inspection indicates that the standard choice of
$\beta=-1$, appropriate for CP-even sum rules, will not be the most
appropriate here as it removes the leading order contribution. 
In general there are two motivated criteria for fixing the
mixing parameter $\beta$: (1) at a local extremum \cite{chung82}; or 
(2) to minimize the effects of the continuum and higher dimensional 
operators \cite{ioffe81,bi83,ioffe83,leinweber}.
We find in this case that extremizing in
the parameter $\beta$ also leads to the unappealing cancelation
of the leading order contribution. Thus the 
most natural procedure appears to be to choose $\beta$ in order to cancel
the subleading infrared logarithm which is ambiguous as a result of 
the required infrared cutoff. This procedure actually 
mimics the effect of the choice
$\beta=-1$ in the sum rule for the nucleon mass. 
We therefore take $\beta=1$, and it is
this choice that we shall now contrast with the phenomenological side 
of the sum rule. It is important to note, however, that
use of the ``lattice'' current with $\beta=0$ will also
produce a numerically similar result. We shall make further comments on this
issue at the end of the section.

On the phenomenological side of the sum rule we have (\ref{phenfinal})
in which the coupling is now interpreted as $\la=\la_1+\beta\la_2$.
After a Borel transform of (\ref{ope}) and (\ref{phenfinal}), and
using $\beta=1$ as discussed above, 
we obtain the sum rule
\be
 \la^2 m_nd_n+AM^2 & = & -\frac{M^4}{32\pi^2}\ov{\theta}
   m_*\langle \ov{q}q\rangle
   e^{m_n^2/M^2}\nonumber\\
   && \!\!\!\!\!\!\!\!\!\!\!\!\!\!\!\!\!\!\!\!\!\!\!
  \times\left[4\ch(4e_d-e_u)-\frac{1}{2M^2}\xi(4e_d+8e_u)\right]
   \label{sumrule}
\ee  

The coupling $\la$ present in (\ref{sumrule}) may be obtained
from the well known sum rules for the tensor structures {\bf 1} and 
$\gsl{q}$ in the CP even sector (see e.g. \cite{leinweber} for a 
recent review). 

To aid in the presentation of the results it 
is convenient to define an additional function $\nu(M^2)$,
\be 
 \nu(M^2) & \equiv & \frac{1}{2\ov{\theta}m_*}
    \left(d_n+\frac{AM^2}{\la^2m_n}\right),
\ee
which is then determined by the right hand side of (\ref{sumrule}).
Inspection of (\ref{sumrule}) suggests that it is not
appropriate in this case to try and remove the unknown parameter $A$, via
differentiation for example. Instead we shall make the 
conventional assumption that the left hand side
is a linear function of $M^2$ (i.e. $A$ is independent of $M^2$), and 
construct two sum rules whose behaviour
will allow us to estimate the slope of this line, and thus the parameter
$A$. In fact this approach will lead to a result which for consistency
will require $A\sim 0$, and thus allow an extraction of the EDM parameter
$d_n$. We now construct these two sum rules as follows:
\begin{itemize}
\item {\bf (a)} Firstly, we extract a numerical value for $\la$ via a direct
analysis of the CP even sum rules. This analysis has been discussed before
and will not be reproduced here (see e.g. \cite{leinweber}). 
One obtains\footnote{\small The particular choice of $\beta$ in the range
$[-1,1]$ is not very important here as the numerical value for 
$\lambda$ is not highly sensitive to this choice.}
\be
   (2\pi)^4\la_a & \sim & 1.05\pm 0.1,
\ee
which leads to a sum rule of the form,
\be
 \nu_a(M^2) & = & -\frac{M^4}{64\pi^2\la_a^2 m_n}\langle \ov{q}q\rangle
   e^{m_n^2/M^2}
  \left[4\ch(4e_d-e_u)-\frac{1}{2M^2}\xi(4e_d+8e_u)\right]
   \label{sumrule:a}
\ee 

\item {\bf (b)} As an alternative,
we extract $\la$ explicitly as a function of $\beta$ 
from the CP-even sum rule for $\gsl{q}$ \cite{leinweber} which we
reproduce here for completeness, 
\be
 (2\pi)^4\la^2e^{-m_n^2/M^2} & = & \frac{5+2\beta+5\beta^2}{64}M^6
   \left[1-e^{-s_1/M^2}\left(\frac{s_1^4}{2M^4}
        +\frac{s_1^2}{M^2}+1\right)\right] \nonumber\\
   & &  \;\;\;\;\;\;\;\;\;
     +\frac{5+2\beta+5\beta^2}{256}bM^2\left(1-e^{-s_2/M^2}\right),
         \label{even}
\ee
where $s_1$ and $s_2$ parametrise the continuum thresholds, while 
$b=(2\pi)^2\langle\frac{\al_s}{\pi}G^2\rangle\sim 0.47\pm 0.2$GeV$^4$, and
we have neglected higher dimensional contributions, and leading-log anomalous
dimension factors. The reason for the omission of the latter is that
they provide a negligible contribution when $M^2$ is small, as will be 
the case here, and furthermore at this scale one has good reason to distrust
the leading log approximation. Therefore, the (estimated) effect of such 
factors will be combined into the overall error estimate.  

Solving (\ref{even}) for $\la_b$, we obtain a new CP-odd sum rule
by substituting the result into (\ref{sumrule}) and setting
$\beta=1$,
\be
 \nu_b(M^2) & = & -\frac{M^4}{64\pi^2m_n}\langle \ov{q}q\rangle
  \frac{\left[4\ch(4e_d-e_u)-\frac{1}{2M^2}\xi(4e_d+8e_u)\right]}
    {4M^6c_1(M^2,s_1)+bM^2c_2(M^2,s_2)}.
   \label{sumrule:b}
\ee
In this expression $c_1$ and $c_2$ are the continuum parametrizations
introduced in (\ref{even}). Throughout we shall assume $s_1=s_2$.

\end{itemize}

We shall now proceed to analyze the sum rules numerically.
Note that our assumption that $\nu(M^2)$ is linear in $M^2$ also requires
that $\la$ is constant in an appropriate range of the Borel
parameter. This point can be checked explicitly in case (b) above.

\begin{figure}
 \includegraphics[width=8cm]{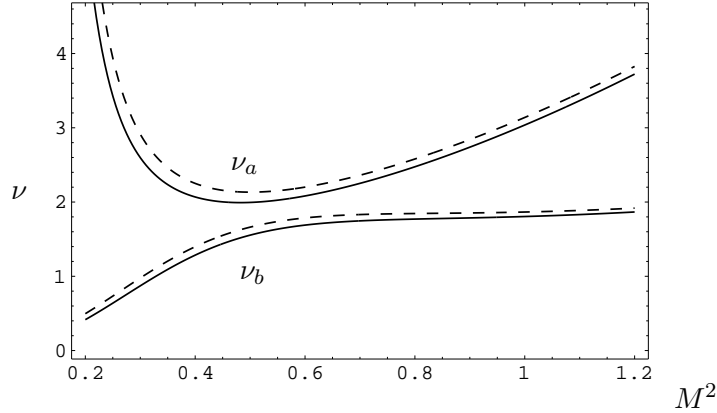}

\vspace{-3cm} \hspace{-9cm} $\nu$

\vspace{-0.9cm} \hspace{-3cm} $\nu_a$

\vspace{0.95cm} \hspace{-2.8cm} $\nu_b$

\vspace{1.2cm} \hspace{9cm} $M^2$

\vspace{0.2cm}

 \caption{\footnotesize The neutron EDM function $\nu(M^2($GeV$^2))$
is plotted according to the sum rules (a) and (b).
The dashed line shows the contribution from the leading order term
only.}
\end{figure}

The two sum rules described above for $\nu_a$ and
$\nu_b$ are plotted in Fig.~4, where the effect of the
higher dimensional terms in (\ref{sumrule}) proportional to
$\xi$ is also displayed. $\nu(M^2)$ is to be interpreted as
a tangent to the curves in Fig.~4.
For numerical calculation we make use of the following
parameter values: For the quark condensate, we take
\be
 \langle 0|\ov{q}q|0\rangle & = & - (0.225\mbox{ GeV})^3,
\ee
while for the condensate susceptibilities, we have the values
\be 
 \ch & = & - 5.7 \pm 0.6 \mbox{ GeV}^{-2}\; \mbox{\protect\cite{chival}},\\
 \xi & = & - 0.74 \pm 0.2\; \mbox{\protect\cite{kw}}.
\ee
Note that $\ch$, which enters at $O(1/M^2)$, since it is dimensionful,
is numerically significantly larger than $\xi$. In extracting $\la$
in case (b) we also set a relatively large continuum threshold
$s_0=(2\mbox{GeV})^2$ for consistency with the CP-odd sum rule in which
this continuum is ignored for reasons discussed earlier.

One observes that both
sum rules have extrema consistent to $\sim 10\%$, suggesting that our
procedure for fixing the parameter $\beta$ is appropriate. Furthermore,
the differing behaviour away from the extrema implies that for consistency
we must assume $A$ to be small. One then finds 
$d_n\sim \nu(M^2\sim 0.5$GeV$^2)$. It is also
interesting that the effective scale is around $M\sim 0.7$GeV which
is well below $m_n$, and should be cause for concern regarding the
convergence of the OPE. Nonetheless, one sees that the corrections
associated with the leading higher dimensional operators are still
quite small. This low scale is also the reason we have ignored
leading-log estimates for the anomalous dimensions as noted
earlier in the context of extracting the coupling $\la$, as their
status is unclear at this scale. A naive application leads to a small
correction that we shall subsume into our error estimate.

Extracting a numerical estimate for $d_n$ from Fig.~2, 
and determining an approximate error arising from:
(1) analysis of the sum rule; (2) the error in $\ch$; and (3) an
estimate of $\pm O(20$\%$)$ for higher dimensional operators and
anomalous dimension factors; we find the result\footnote{(04/2005) v4: This updated expression corrects
an overall factor of two error in previous versions.}
\be
 d_{n} &=& (1.0 \pm 0.4)e\ov{\theta}\frac{m_*}{({\rm 500 MeV})^2}
      = (3.6\pm 1.4)\cdot 10^{-3}
    e \ov{\theta}\frac{f^2_\pi m_{\pi}^2}{(100 {\rm MeV})^5} 
    \frac{m_um_d}{(m_u+m_d)^2},
\label{final}
\ee
for the neutron EDM, for which the dominant contribution naturally arises
from $\ch$. Comparison with the result of Ref. \cite{CDVW} 
indicates rather good agreement in magnitude, due essentially to the 
low effective mass scale $M\sim 700$MeV.
We also obtain $d_n$ of the same sign if one assumes no significant 
corrections to the logarithm in \cite{CDVW}
\footnote{\small The sign of $d_n(\theta)$ 
in \cite{KL} differs simply because of 
an overall minus sign in the initial definition of the $\theta$--term.}. 
The relation between the
calculation presented here and the chiral logarithm 
estimate will be discussed in more detail in the next section.

However, before turning to this comparison, we would like to comment further
on the specific choice of the current, i.e. the choice of $\beta$.
As mentioned earlier, a conventional choice for CP-even sum rules
is that advocated by Ioffe \cite{ioffe83}: $\beta=-1$. 
However, even in these channels, standard minimal sensitivity 
arguments, (as used for example in the context of renormalization 
scheme dependence \cite{stevenson}), lead instead to a choice 
of $\beta\sim -0.2$ \cite{chung82}. Furthermore, we have found 
that an apparently optimal choice for the chirally invariant CP-odd 
channel is instead $\beta=1$. 

This variation in $\beta$ might be
interpreted as a sign of inherent uncertainties due to the effect
of excited states or higher order condensates. However, we would 
like to point out that this need not be the case if these differences 
are observed for different physical observables, as is the case here. 
In particular, while one expects that the optimization of $\beta$ with
respect to minimizing contamination from excited states may be
universal, Ioffe \cite{ioffe83} has emphasized that this aim conflicts
with the need to minimize the uncertainties due to higher dimensional
operators. The latter point is quite ``observable dependent'', and thus
the best compromise choice for $\beta$ may differ for different
observables. In particular, as pointed out in \cite{leinweber}, the
points $\beta\sim -1$ and $+1$ are not distinguished purely on the basis
of removing contamination by excited states.

While we find such arguments for the overall consistency of the
current with $\beta=1$ compelling enough, we have also 
explicitly tested the $\beta$--dependence of our result. To do this,
it is necessary to consider the effect of the logarithmic terms
in (\ref{ope}). Cutting the logs at $\mu_{IR}\sim \La_{QCD}$, we then find
that the numerical value for $d_n$ does not vary that dramatically
with $\beta$. Indeed for $\beta\sim 0$, we find a result which
is numerically within $\sim 30$\% of the result obtained at 
$\beta=1$.  

\section{Relation With Estimates from Chiral Perturbation Theory}

In this section, we shall consider the relation between the value
for $d_n$ we have extracted using QCD sum rules, and the well known
chiral loop calculation \cite{CDVW}. Within the sum rules approach,
use of the chiral anomaly made it quite explicit that the result would
necessarily vanish in the limit $m_{\et}\rightarrow m_{\pi}$.
To compare this with the chiral result, we first need to 
observe how the latter also satisfies this necessary consistency
condition. As far as we are aware, this result 
has not appeared in the literature and we present a derivation 
of the expected prefactor $(1-m_{\pi}^2/m_{\et}^2)$ in the next
subsection.

\begin{figure}
 \includegraphics[width=8cm]{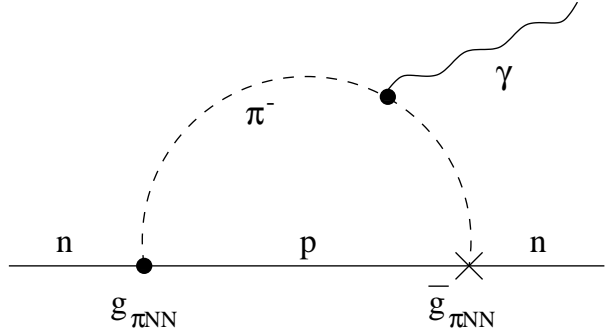}

 \caption{\footnotesize Leading contribution as $m_{\pi}\rightarrow 0$ to 
      $d_n$ in the chiral loop approach \protect\cite{CDVW}. 
      There is also a second
      diagram with the pion-nucleon vertices interchanged.}
\end{figure}

Prior to addressing this issue, we first recall the main points of the
chiral approach used in \cite{CDVW}. A chiral rotation is used to recast
the $\theta$--term into a CP violating quark mass, 
$\de {\cal L}_{CP} = i\theta m_* \sum_f \ov{q}_f \ga_5 q_f$. This induces
an effective CP-odd interaction of pions with nucleons in terms of a small 
correction to the standard $\gamma_5$-interaction,
\be
{\cal L}_{\pi NN}= \pi^a \bar N \tau^a( 
ig_{\pi NN} \gamma_5 +\bar g_{\pi NN})N.
\ee
The CP-odd pion-nucleon
coupling $\ov{g}_{\pi NN}$ induced by $\theta$ may be obtained by the 
PCAC reduction of the pion  
\be
\bar g_{\pi NN}\bar N\tau^a N = - \fr{i}{2f_\pi}
\langle N|[\de {\cal L}_{CP}, \bar q \tau_a q ] |N\rangle= 
\fr{\theta m_*}{f_\pi}
\langle N|\bar q \tau^a q |N\rangle .
\label{gbar}
\ee
The remaining matrix element can then be expressed in terms of the
mass splittings in the baryon octet using $SU(3)$ flavour symmetry. 
Note that the expression (\ref{gbar}) is valid only for soft pions with 
momenta smaller than the characteristic hadronic scale of 1 GeV. 

The crucial observation made in \cite{CDVW}
is that the electric dipole moment of the neutron, induced by 
the source $\de {\cal L}_{CP}$, can be estimated using 
the singular behaviour of the chiral loop shown in Fig.~5. 
The loop integral has a logarithmic infrared divergence as 
$m_{\pi}\rightarrow 0$ and one then obtains 
the estimate \cite{CDVW}
\be
 d_n^{\rm cpt} & = & eg_{\pi NN} \ov g_{\pi NN} \frac{1}{4\pi^2m_N}\ln 
        \frac{\La}{m_{\pi}},
 \label{cloop}
\ee 
where the cutoff represents the scale at which chiral perturbation 
theory breaks down. It is reasonable to assume that 
$\La\sim m_{\rh}$ or $m_n$. After substituting the numerical expression
for $\bar g_{\pi NN}$ we arrive at the result reproduced in 
(\ref{eq:log}).

\subsection{$U_A(1)$-properties of the EDM in chiral loop calculations}

The independence of the original chiral loop result (\ref{cloop})
from $m_\eta$ has been the source of some controversy, and continued work,
in the literature (see e.g. \cite{AH}).
Before we compare our results with that of Ref \cite{CDVW}, we wish to 
point out the particular mechanism, ignored in the 
discussion above, which sets the EDM to zero when $U_A(1)$
symmetry is restored. This cancelation should, of course, be very 
similar to the vanishing of the CP-odd amplitude for $\eta\rightarrow \pi\pi$,
demonstrated in \cite{svz}.

Within the chiral loop approach reviewed above, the EDM depends explicitly
on the CP-odd pion--nucleon coupling  $\bar g_{\pi NN}$ induced by $\theta$.
We shall now prove that this coupling is 
proportional to the factor $(1-m_\pi^2/m_\eta^2)$ for the case of 
two flavors, taking $m_u=m_d\equiv m_q$ for simplicity. 

Recall that 
in the calculation of \cite{CDVW}, the $\theta$--term is written in the 
form of a singlet combination of quark $\gamma_5$--mass terms. 
Since the $\eta$--meson has the same 
quantum numbers, this combination can produce $\eta$ from vacuum with an 
amplitude proportional to $f_\pi^{-1}m_q\langle 0| \bar qq|0\rangle$. Thus, 
the calculation of $\bar g_{\pi NN}$ should account for the 
additional contributions related to $\eta$ being produced from the vacuum 
and then reabsorbed by the nucleon, subsequently 
producing the soft pion (see Fig 6).

\begin{figure}
 \includegraphics[width=13cm]{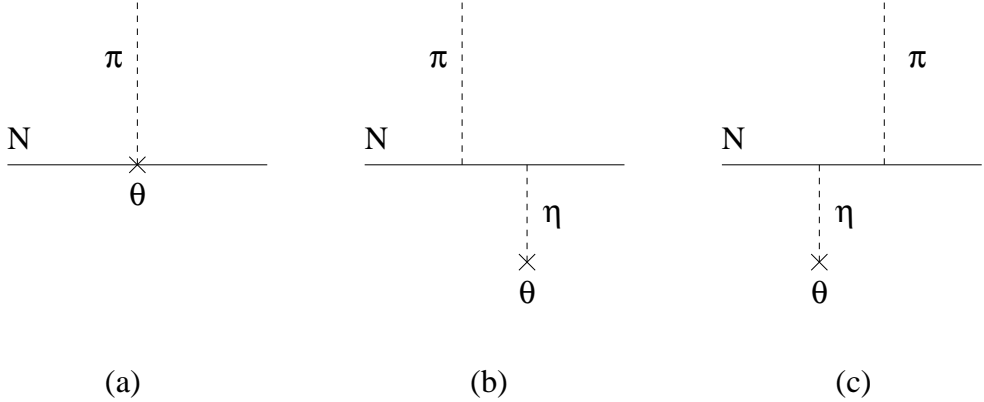}

 \caption{\footnotesize Different contributions to the $\bar g_{\pi NN}$ coupling. 
The leading diagram (a) is identically cancelled by (b) and (c) in the limit
$m_\eta \rightarrow m_\pi$.}
\end{figure}

In the absence of any physical effect from the anomaly, 
the mass of $\eta$ should vanish in the chiral limit. In this case 
the flavor--singlet field $\eta$ can be treated exactly as a pion field. 
Therefore, the amplitude for the low energy scattering 
$\pi N\rightarrow\eta N$ can be related to the nucleon sigma term 
in the triplet channel,
\be
M_{\pi^a N\rightarrow\eta N} \sim m_q \langle N|\bar q \tau^a q|N \rangle,
\ee
which is exactly the structure appearing in the calculation of the CP-odd 
pion--nucleon coupling constant $\bar g_{\pi NN}$, eq. (\ref{gbar}).

Summing different contributions we obtain
\be
\bar g_{\pi NN}\bar N\tau^a N =  \fr{\theta m_q}{2 f_\pi}
\langle N|\bar q \tau^a q |N\rangle \left(1 - \fr{1}{f_\pi^2}
\fr{2m_q\langle \bar qq \rangle}{m_\eta^2}\right)= \fr{\theta m_q}{2 f_\pi}
\langle N|\bar q \tau^a q |N\rangle\left(1-\fr{m_\pi^2}{m_\eta^2}\right),
\label{abc}
\ee
with the anticipated dependence on $m_\eta^2$. 

Now the connection to the $U_A(1)$ problem and the vanishing of 
$\bar g_{\pi NN}$ in the limit $m_\eta \rightarrow m_\pi$ become explicit. 
In real life, due to the physical effect of the anomaly, 
$m_\eta$ remains finite in the chiral limit, whereas $m_\pi\rightarrow 0$, 
and the correction from diagrams (b) and (c) is negligible as it is 
second order in the light quark masses. Thus numerical extraction
of bounds on $\theta$ using this result, as in \cite{CDVW}, are 
essentially unaffected by this correction.
Nevertheless we find this exercise rather 
instructive, noting that the coupling 
$\bar g_{\pi NN}(\theta)$ may also be used for the 
extraction of the limit on $\theta$ from the mercury 
EDM experiment \cite{mEDM}. 

As a small digression,
it is also worth pointing out that in \cite{AH}, where the issue of
$U_A(1)$ restoration was also addressed, it was suggested that
contributions to $d_n(\theta)$ can be obtained within the chiral
approach which are directly proportional to the 
anomalous magnetic moment of neutron. This seems highly unlikely since
$\mu$ and $d$ are generated at quite different energy scales. 
The one-loop diagram 
contributing to $d$ receives contributions from momentum scales 
$m_\pi^2\ll p^2\ll 1 {\rm GeV}^2$, whereas similar diagrams for 
$\mu$ diverge quadratically and thus are saturated by momenta $\sim$ 1 GeV
where the chiral description breaks down.

\subsection{Relation With the EDM From Chiral Perturbation Theory}

Given the numerical similarity between the result we have obtained
for $d_n(\theta)$ and the estimates based on the dominance of 
the chiral logarithm, $\ln m_{\pi}$, it is natural to ask whether or not
this logarithm is hidden somewhere in the OPE analysis.

Such a suggestion is necessarily speculative, as the calculations are
performed using different dynamical degrees of freedom, and are in principle
valid at quite different external momentum scales. Nonetheless,
at first sight, one may be tempted to identify the chiral
log term explicitly with the subleading infrared log--terms obtained 
in (\ref{ope}). The full momentum dependence of this term is, 
however, $\ln (Q^2/\mu^2)/Q^2$, and thus the singular behaviour is 
instead determined by $1/Q^2$. Nevertheless,
one might suggest that this naive extrapolation to the chiral regime
is not appropriate, and the power-like term gets softened, while the
logarithm remains. 
The most obvious log--term of this kind arises from diagram (b) in 
Fig.~2. Assuming the top two quark lines are soft, as appropriate for the
condensate, we can interpret this as a soft pion, and by adding
two spectator quarks, we get something analogous to the chiral loop
diagram contributing in the approach of \cite{CDVW}.
Here, CP--violation arises from a particular 
4-fermion vertex proportional to $\theta$.

While such heuristic relations are sometimes possible when the momentum
dependence is easily equated on both sides, 
we would now like to argue that, at least at large $N_c$, 
the chiral logarithm is actually subleading with respect to the
leading order term in the OPE. While this need not be important
numerically for $N_c=3$, it provides a convenient parametric
distinction between the results obtained using sum rules and chiral
perturbation theory. 

At large $N_c$, the $U_A(1)$ restoration factor
for the chiral expression derived in the previous subsection 
becomes important. 
In particular, as is clear from the discussion above, 
the result of calculation, either with the use of chiral 
techniques or QCD sum rules, is proportional to $(1-m_{\pi}^2/m_{\et}^2)$, 
and thus the EDM has a natural parametric dependence on $\theta$ of
the form $\theta/N_c$. 

However, the main distinction between the sum rules and chiral
loop calculations is that the quark diagrams used within the sum rules
calculation are effectively tree-level. Thus one expects the effect
to be parametrically enhanced relative to the chiral loop.
Indeed, this follows from  an additional suppression factor of
$1/f_\pi^2$, which is $O(1/N_c)$, and which is absent in the OPE calculation.
The corresponding factor has the form $\langle \ov{q}q\rangle/m_n$,
which is $O(N_c/N_c\sim 1)$. Summarizing, we obtain the following 
behaviour at large  $N_c$ in these two 
cases,
\begin{eqnarray}
\mbox{QCD-SR}~~~~ d_n &\propto&\fr{\theta}{N_c}
\fr{\langle 0|\bar qq |0\rangle}{m_n}\sim O(\theta/N_c)\nonumber\\
\mbox{Chiral Loop}~~~~ d_n&\propto& \fr{\theta}{N_c}
\fr{\langle N|\bar uu -\bar dd |N\rangle}{f_\pi^2}\sim O(\theta/N_c^2)
\end{eqnarray}
where we have only kept track of the relative $N_c$--dependent 
factors\footnote{An additional $N_c$ enhancement, due to 
the scaling of $g_A\sim N_c$ has been discussed in relation to Skyrme
model calculations (see e.g. \cite{skyrme}). This effect should be
generic to both approaches, and we have ignored such overall contributions.} 
It is interesting to note that the isovector matrix element 
$\langle N|\bar uu -\bar dd |N\rangle$ contributing to the chiral
result does not grow with $N_c$,
whereas both $\langle N|\bar uu|N\rangle$ and $\langle N|\bar dd|N\rangle$
are proportional to $N_c$. This is because the neutron in the large
$N_c$ limit has $(N_c-1)/2$ $u$-quarks and $(N_c+1)/2$ $d$-quarks.
 
In summary, while there appears to be at least a qualitative mechanism
for mapping some of the OPE diagrams to those which would produce
a chiral logarithm, the behaviour at large $N_c$ is of the form
$(c+\ln m_{\pi}/N_c)$. Of course, we reiterate that
physically there need be no suppression for $N_c=3$.

\section{Discussion}

The calculation of $d_n(\theta)$ via QCD sum rules has produced a
numerical result very close to the estimates obtained using different 
techniques. Indeed the chiral loop estimate, QCD sum rule calculations, 
and even the naive quark model \cite{baluni}, all agree and predict 
essentially the same EDM to within 
a factor of 2-3. It is interesting to note that the power counting 
procedure in the chiral theory, combined with the quark model for 
nucleons known as naive dimensional analysis \cite{GM}, also 
produces a similar estimate. 

The situation is apparently very
different in the case of dimension 5 and 6 CP-odd operators, especially 
for color EDMs. Different methods produce results varying within
more than one order of magnitude. Moreover, none of the existing calculations 
is capable of answering the question of which combination of color EDMs  
of $u$, $d$ and $s$ quarks in fact enters into the single observable
$d_n$. We believe that the approach developed in this paper can be 
used for the calculation of the neutron EDM induced by
these dimension 5 and 6 CP-odd operators, and such a 
calculation will clarify this issue. A technical outcome of the work presented
here is that the analysis can be done in the 
chirally invariant channel, where the OPE is directly proportional to the 
electric dipole moment. 

There is an additional incentive for using QCD sum 
rules for the calculation of $d_n$, as opposed to any other method. 
In the case of dim=5 operators, QCD sum rules would normally produce 
$d_n$ in the form of a linear combination of terms like
$d_q\langle \bar qq \rangle$, 
where $d_q$ is an EDM or a color EDM of a light quark. 
In many models, including the MSSM and variants thereof, 
$d_q\sim m_q$ which would allow the removal of much of the uncertainty 
related to the imprecise knowledge of light quark masses. Indeed, when 
multiplied by the value of the quark condensate, any linear
combination of $d_q$ can be expressed in terms of $m_\pi^2f_\pi^2$ times
a function depending only on the ratio of light quark masses. The latter is 
known to much better accuracy than the quark masses themselves and does not 
depend on the normalization scale. 

In conclusion, we have presented a QCD sum rules calculation of 
the $\theta$-induced neutron EDM. This result is explicitly tied
to a set of vacuum correlators which are non-vanishing only in the absence
of a U(1) ``Goldstone boson''. The use of QCD sum rules in the chirally 
invariant channel allowed us to unambiguously extract $d_n(\theta)$, and 
independence of the answer from any particular representation of the theta
term (\ref{general}) was checked explicitly.

Combining our result with the recently improved experimental bound on $d_n$
\cite{nEDM1} we derive the limit on theta:
\be
|\bar \theta|< 3\times 10^{-10},
\ee
which is quite close to previous bounds, and actually somewhat less
constraining as our result for $d_n(\theta)$ is slightly lower
than the corresponding result obtained within chiral perturbation theory
with the cutoff at $m_\rho$. Numerically, to a large extent, our 
result is proportional to $\chi$, the electromagnetic 
susceptibility of the vacuum.

\vspace{0.1in}
{\bf Acknowledgments:}
We would like to thank I. Khriplovich, W. Marciano, M. Shifman, N. Uraltsev, and 
especially A. Vainshtein for many valuable discussions
and comments.  This work was supported in part by
the Department of Energy under Grant No. DE-FG02-94ER40823.

\bibliographystyle{prsty}

\end{document}